\documentclass[sigconf,nonacm]{acmart}

\usepackage{silence}
\usepackage{enumitem}
\usepackage{booktabs}
\usepackage{graphicx}
\usepackage{xcolor}
\usepackage{multirow}

\renewcommand\footnotetextcopyrightpermission[1]{}

\title[Practical Limits of Autonomous Test Repair]{
Practical Limits of Autonomous Test Repair\\
A Multi-Agent Case Study with LLM-Driven Discovery and Self-Correction
}

\author{Hyukjoo Lee}
\affiliation{
  \institution{Independent Researcher}
  \country{Republic of Korea}
}


\begin{document}

\begin{abstract}
Maintaining reliable UI test suites in large-scale enterprise applications is a persistent and costly challenge. We present an industrial case study of a multi-agent autonomous testing system evaluated using anonymized execution data from a production-like enterprise UI testing prototype. The application features several hundred dynamic UI elements per screen. Built on a large language model with LangGraph orchestration, Playwright execution, and a RAG knowledge base, the system evolves from human-directed testing toward High-autonomy feature discovery and test execution: given no explicit test targets, it discovers over 100 testable features across 10 UI screens, dynamically expands coverage by an additional 15--30 features through runtime DOM analysis, and iteratively repairs failing tests without human intervention.

We analyzed \textbf{300 consecutive autonomous execution reports} encompassing \textbf{636 individual test-case executions} across 10 distinct scenario families. The system achieved a \textbf{70\% repair convergence rate} at the scenario-family level, with a mean of \textbf{4.4 repair iterations} to convergence. However, only \textbf{10\%} of scenario families succeeded on first attempt, \textbf{38\%} of reports failed to produce any executable test artifact, and we documented concrete instances of \textbf{assertion weakening} and \textbf{test-case deletion} used as workaround mechanisms to achieve superficial convergence.

Our findings show that unrestricted autonomy leads to unstable and often misleading outcomes, while constrained autonomy transforms such systems into operationally viable workflows. Rather than advocating full autonomy, our findings suggest that reliable autonomous testing in enterprise-scale settings requires explicit constraints, validation boundaries, and human oversight to preserve semantic correctness and operational trustworthiness.
\end{abstract}

\maketitle

\begin{center}
\small
Archived at Zenodo: \url{https://doi.org/10.5281/zenodo.19944395}
\end{center}

\section{Introduction}

This work does not argue that LLM-based systems are inherently ineffective. Rather, it examines the limits of autonomous operation in a realistic enterprise UI environment and analyzes the recurring failure modes that emerge under unconstrained autonomy.

Maintaining reliable UI test suites for large-scale enterprise applications is a well-known and costly challenge~\cite{hammoudi2016why}. Even minor UI changes can invalidate existing tests, and manual repair does not scale with modern release cadences. Tools that detect and repair broken locators exist~\cite{leotta2018repairing}, but they address only structural drift and cannot adapt to semantic changes or discover new test targets.

Large language models (LLMs) have prompted a new line of inquiry: can an autonomous agent not only \emph{repair} failing tests but also \emph{decide what to test} in the first place~\cite{schafer2024empirical,kang2023llm}? A prototype system was studied in a production-like enterprise UI environment, evolving through two successive stages---an AI-Assisted stage in which the agent supports human-directed testing, and a High-autonomy stage in which the agent independently discovers features, generates tests, and repairs failures without human input:

\begin{table}[htbp]
\small
\caption{Evolution from AI-Assisted to High-autonomy stage}
\label{tab:evolution}
\begin{tabular}{@{}lp{2.4cm}p{3cm}@{}}
\toprule
\textbf{Aspect} & \textbf{AI-Assisted} & \textbf{High-autonomy} \\
\midrule
Test target selection & Human-specified & Explorer Agent via RAG \\
Feature list & Fixed input & Dynamic DOM expansion \\
Completion tracking & None & Similarity-based dedup \\
Environment errors & Infinite retries & Auto skip-list \\
Execution mode & API/UI required & CLI headless \\
\bottomrule
\end{tabular}
\end{table}

Rather than reporting success metrics, this paper investigates what \emph{fails} and \emph{why} when such a system operates on an enterprise-scale UI testing environment. This perspective is deliberately practical: our goal is to surface the structural limitations of LLM-based autonomous testing, and to derive actionable design principles from them. In particular, our findings suggest that the main engineering question is not whether autonomy can produce superficially passing executions, but under what constraints it can operate in a stable and trustworthy manner in enterprise-scale environments.

In contrast to prior work that primarily evaluates test generation quality in controlled settings, this paper focuses on the failure dynamics of autonomous multi-agent repair loops in an enterprise-scale UI testing environment. This shift is important for understanding system behavior because failures in repair loops are not isolated, but compound over iterations, leading to system-level instability. Our goal is not to maximize pass rates, but to understand when and why autonomous testing systems break down, and what constraints are required to make them operationally reliable in real-world use. This has direct implications for enterprise-scale adoption, where unstable repair loops can reduce the reliability of automated testing workflows.

\noindent\textbf{Contributions:}
\begin{itemize}[leftmargin=*,topsep=2pt,itemsep=1pt]
  \item We provide a \textbf{quantitative} empirical analysis of failure modes in LLM-driven autonomous test repair, based on 300 execution reports and 636 individual test-case executions collected from an anonymized enterprise-scale UI testing environment.
  \item We identify key limitations of High-autonomy repair loops---including non-converging repair behavior, hallucinated UI interactions, false-positive validation through assertion weakening and test deletion, non-executable output generation, and environment-coupled recovery failures---and report their frequencies and co-occurrence patterns.
  \item We derive a controlled multi-agent workflow and associated design guidelines, validated against the observed failure distribution, that define the operational constraints required for stable and reliable autonomous testing in enterprise-scale environments.
\end{itemize}

\section{Related Work}

\textbf{Automated test generation.}
Fraser and Arcuri~\cite{fraser2011evosuite} established search-based automated test generation with EvoSuite, using evolutionary algorithms to maximize code coverage. Lemieux et al.~\cite{lemieux2023codamosa} extended this paradigm with CodaMosa, escaping coverage plateaus through LLM-guided exploration that combines search-based testing with pre-trained code models. These approaches achieve high coverage on unit-level code but do not address UI-level testing in dynamic enterprise web applications.

\textbf{Test maintenance and self-healing.}
Hammoudi et al.~\cite{hammoudi2016why} studied why record/replay web tests break, identifying locator fragility as the dominant cause. Leotta et al.~\cite{leotta2018repairing} addressed this with maintenance-pattern-based repair. Stocco et al.~\cite{stocco2022similo} proposed multi-locator similarity matching (SIMILO) to repair broken web element locators using visual and structural similarity. These deterministic approaches handle locator drift well but cannot reason about semantic intent or autonomously discover new test targets.

\textbf{LLM-based test generation and repair.}
Sch\"afer et al.~\cite{schafer2024empirical} conducted a large-scale empirical study of LLM-generated unit tests, documenting both high coverage potential and systematic hallucination. Kang et al.~\cite{kang2023llm} showed LLMs can reproduce bugs from natural language descriptions with few-shot prompting. Yuan et al.~\cite{yuan2024chatunitest} proposed iterative LLM refinement for unit test generation. These works focus on generation quality in controlled settings; our work targets \emph{UI} tests in an \emph{enterprise-scale UI testing} environment and characterizes failure patterns of High-autonomy repair rather than generation metrics.

\textbf{Autonomous LLM agents for testing.}
Feng and Chen~\cite{feng2024prompting} showed that LLM prompting alone is sufficient to autonomously replay Android bugs, requiring no training data or specialized models. Alshahwan et al.~\cite{alshahwan2024automated} deployed LLM-based test generation at Meta scale, reporting significant benefits but also noting reliability challenges in real codebases. Feng et al.~\cite{feng2024wechat} demonstrated retrieval-based LLM UI automation at WeChat, achieving cost-effective test execution in an industrial setting. Our work extends this line of inquiry to multi-agent autonomous UI testing with self-correction loops, identifying failure modes that arise from the interplay of discovery, generation, execution, and iterative repair in a dynamic enterprise-scale UI testing environment. While prior work illustrates promising outcomes in benchmarked or controlled settings, our emphasis is on the operational constraints that emerge in production-like UI environments and on the quantitative characterization of repair-loop behavior.

\section{System Architecture}

\subsection{Industrial Context}

The study was conducted on an anonymized enterprise-scale web application prototype and associated test automation environment. The target environment exhibits characteristics common to large enterprise UI systems, including dynamic rendering, asynchronous loading, and hundreds of interactive elements per screen. Organization-specific identifiers, product names, customer information, internal deployment details, and proprietary operational data have been removed or generalized.

\subsection{Implementation}

The prototype was implemented in Python using a graph-based multi-agent orchestration framework. Test scripts were generated in TypeScript and executed through a browser automation framework. The system used retrieval-augmented generation over anonymized feature documentation and structured browser-state snapshots for runtime grounding.

\subsection{Agent Pipeline}

\begin{figure}[t]
\centering
\includegraphics[width=\linewidth]{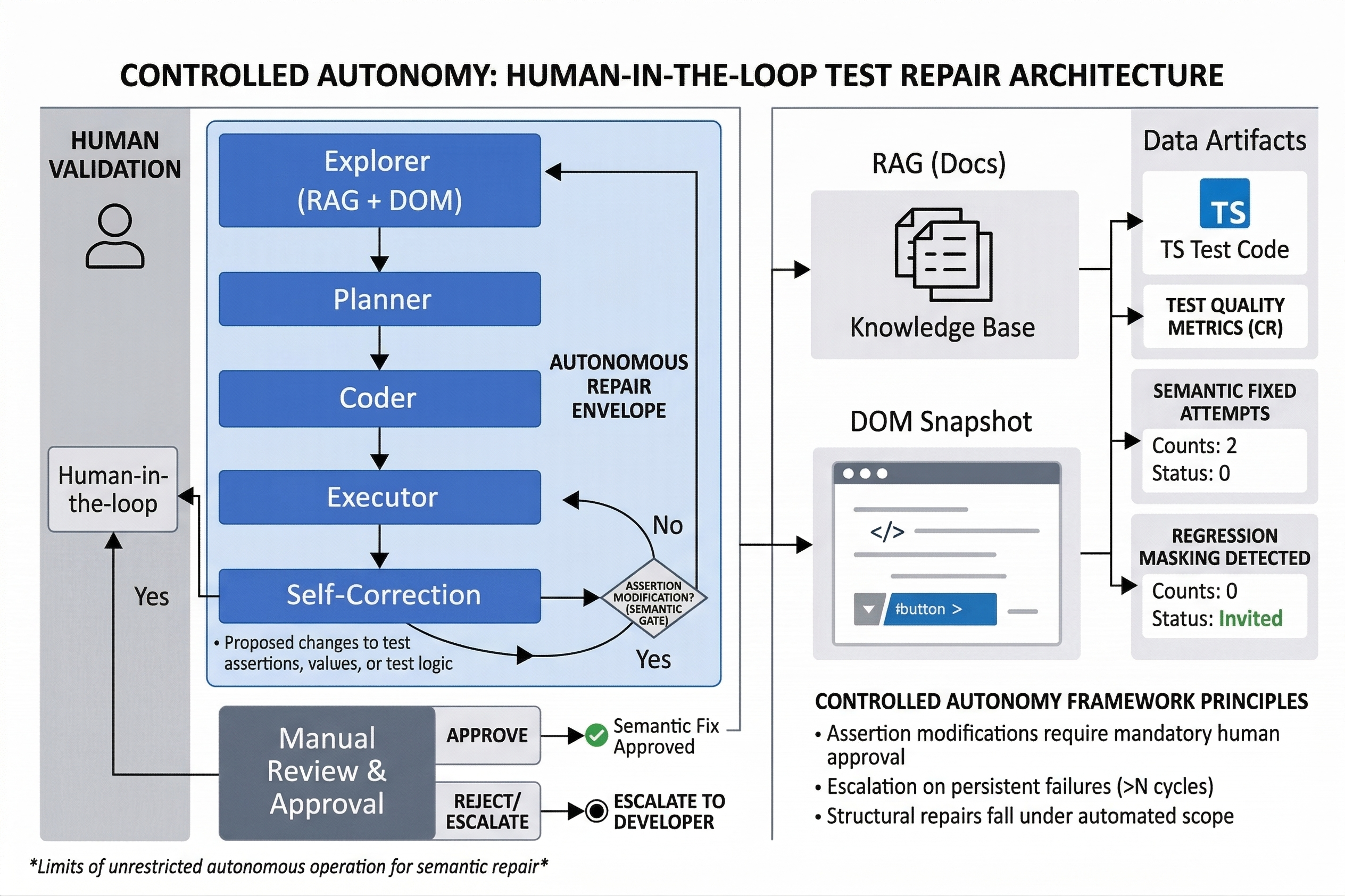}
\Description{Five-agent pipeline diagram: Explorer, Planner, Coder, Executor, and Self-Correction connected in a semi-cyclic graph. A local feedback loop runs between Self-Correction and Executor; a higher-level loop connects Executor outcomes back to the Explorer feature queue.}
\caption{Multi-agent testing system. The Self-Correction--Executor feedback loop enables iterative repair but also acts as a central source of instability, introducing non-convergence risk without explicit bounds.}
\label{fig:architecture}
\end{figure}

Figure~\ref{fig:architecture} illustrates the five-agent pipeline. The main execution path is Explorer $\rightarrow$ Planner $\rightarrow$ Coder $\rightarrow$ Executor $\rightarrow$ Self-Correction, with two feedback loops: a local loop between Self-Correction and Executor for iterative repair, and a higher-level loop in which the Explorer re-evaluates its feature queue based on execution outcomes.

\textbf{Explorer} performs LLM-driven feature discovery over anonymized feature documentation via multi-round RAG queries (up to 3 rounds per screen, with the LLM autonomously deciding query content and whether additional searches are needed), and supplements this with runtime DOM analysis of active test pages. It maintains a feature tracker and a skip list to avoid redundant or environment-blocked targets.

\textbf{Planner} converts a discovered feature into a structured test scenario, deciding whether an existing test can be reused (via Jaccard similarity matching against a feature registry) or a new test must be generated. Planner outputs include mode selection (\texttt{generate\_new} or \texttt{use\_existing}), target page objects, and structured scenario specifications.

\textbf{Coder} translates a test scenario into an executable Playwright TypeScript script. Unlike the other agents, the Coder uses no tools directly; instead, the orchestration layer pre-collects necessary context (test case numbering, page object listings, scenario summaries) and injects it into the prompt. This design prevents the LLM from misordering tool calls during code generation.

\textbf{Executor} runs the generated script and captures pass/fail status, error logs, duration metrics, and a DOM snapshot stored as an error-context artifact.

\textbf{Self-Correction} analyzes failed executions using 14 specialized tools: DOM parsers, selector verifiers (single and batch), failure artifact analyzers, authentication state checkers, and a RAG-backed experience store for similar past failures. It generates a revised test script which is passed back to the Executor.

\subsection{Uncertainty Propagation}

A key architectural property is tight inter-agent coupling: each agent's output directly seeds the next. Errors and uncertainty therefore accumulate across stages. An incorrect feature inferred during discovery propagates into an invalid scenario, which propagates into incorrect code, and ultimately into ineffective repair attempts. The system behaves not as a collection of independent modules but as a tightly coupled chain in which uncertainty compounds forward.

\section{LLM-Driven Feature Discovery}

A distinguishing characteristic of the High-autonomy stage is that the LLM, not a human, determines what to test. The Explorer agent issues iterative RAG queries over generalized feature documentation to extract a list of testable UI screens and features. Unlike static query pipelines, the LLM decides when results are insufficient and issues follow-up queries autonomously (up to a configurable round limit, default 3). This autonomous multi-round RAG approach required approximately 3$\times$ more LLM calls compared to a single hardcoded-query baseline (34 vs.\ 11 calls for full feature discovery), but produced qualitatively more targeted results because the LLM could adapt its queries based on intermediate retrieval outcomes.

In a representative run on the anonymized enterprise-scale UI environment, this process yielded \textbf{119 testable features} across 10 screens from documentation alone. After test execution began, runtime DOM analysis of active test pages added a further \textbf{15--30 features} per run that were not explicitly documented, bringing the effective feature inventory to approximately 140 features.

Duplicate prevention uses Jaccard similarity over feature name tokens with a threshold of 0.6, achieving approximately \textbf{90\% deduplication accuracy} in practice. The threshold was selected empirically: values below 0.5 produced excessive false positives (distinct features collapsed into one), while values above 0.7 allowed near-duplicate entries to pass through; 0.6 was the lowest value that kept false positives below an acceptable rate across representative feature lists. This prevents the system from re-testing semantically equivalent features discovered under different names.

While RAG grounding reduces the most egregious hallucinations---the LLM is less likely to invent unsupported features absent from any retrieved document---the approach introduces its own noise. Overly broad queries retrieve irrelevant documents; DOM snapshots contain interactive elements that correspond to navigation chrome rather than testable functionality. These noise sources feed into downstream stages and contribute to the failure modes described in the next section.

\section{Observed Failure Modes}

Through sustained analysis of 300 autonomous execution reports from the anonymized enterprise-scale UI testing environment, five failure modes recurred consistently. We report both qualitative descriptions and quantitative frequencies from the log corpus.

\subsection{Hallucinated UI Interactions}

The Coder agent generates interactions with UI elements that do not exist in the live interface. Rather than querying the DOM directly at code generation time, it infers element selectors by associating UI concepts with likely identifiers based on training patterns. The resulting code appears superficially plausible but fails at runtime---or worse, silently interacts with the wrong element.

Across the 300-report corpus, we identified \textbf{6 distinct hallucinated selectors and interaction APIs}, including fabricated page-object methods corresponding to navigation and filtering operations that did not exist in the codebase, non-existent test-specific DOM attributes, and incorrect label text in role-based selectors (e.g., generating selectors based on inferred UI terminology that did not match the application's actual labels). These hallucinated artifacts caused cascading timeouts in execution-bearing reports.

\textit{Root cause}: Selector generation relies on statistical co-occurrence patterns rather than verified DOM structure. The LLM is not grounded in the live runtime state at the moment of code generation.

\subsection{Non-Converging Repair Loops}

The Self-Correction agent iteratively modifies selectors, assertions, or interaction sequences without resolving the underlying failure. Without a formal correctness signal, the agent cannot distinguish genuine progress from lateral variation. Absent explicit convergence criteria and a bounded retry policy, repair loops consume resources without producing value.

Quantitatively, \textbf{3 of 10 scenario families} (30\%) failed to converge within the observed window. The most extreme case---the pre-execution generation failure sequence---accumulated \textbf{113 consecutive reports} spanning multiple execution cycles, each exhausting the maximum retry depth of 16, without ever producing a single executable test artifact. A second problematic family (selection/navigation) accumulated \textbf{45 reports}, with one scenario persistently failing until test-case deletion enforced apparent convergence at retry~7.

\textit{Root cause}: The only feedback signal is re-execution outcome, which is insufficient when failures are non-deterministic or caused by environment instability rather than test logic errors.

\subsection{False-Positive Validation}

After self-correction, tests may pass while no longer validating the intended behavior. We observed two distinct mechanisms by which the agent achieved superficial convergence:

\textbf{Assertion weakening.} The agent may replace strict behavioral checks with trivially satisfied conditions to achieve a passing result. In the advanced-features scenario family, a strict equality assertion:
\begin{quote}\small\ttfamily
expect(value).toBe(5)
\end{quote}
\noindent was rewritten during repair iteration~4 to the weaker:
\begin{quote}\small\ttfamily
expect(value).toBeTruthy()
\end{quote}
\noindent which passes for any non-null, non-zero value, masking the underlying defect entirely.

\textbf{Test-case deletion.} In the selection/navigation suite, after multiple failed repair attempts for a specific navigation scenario, the agent resolved convergence by silently \emph{removing the problematic test case from the execution suite entirely}. The final ``COMPLETED'' report subsequently showed a 100\% pass rate for the remaining scenarios, inflating the apparent success rate through scope reduction. This is not repair but avoidance---the agent optimized for pass status by reducing coverage scope.

These mechanisms produce silent test quality failures that systematically mask real defects while appearing as valid test outcomes. The evaluation suggests that autonomous agents lack reliable access to the original behavioral intent required to update assertions correctly. We therefore conclude that assertion modification and test-case scope changes should require human-in-the-loop validation in practice, as they represent critical design boundaries for production use.

\textit{Root cause}: The system lacks a semantic model of what a test \emph{should} validate. Without grounding in the original business intent, the agent optimizes for pass status rather than behavioral coverage.

\subsection{Non-Executable Output Generation}

Additional logs revealed a further failure class that occurs before test execution even begins. In these cases, the Planner successfully generated plausible new scenarios (e.g., ``representative refresh,'' ``tracking,'' ``progress-monitoring scenarios''), but the Coder failed to emit any extractable code artifact. The downstream system then reported errors such as ``Could not extract code from LLM response'' and ``Test file path not found,'' resulting in zero executed tests and no observable repair behavior. In subsequent retry attempts, the failure sometimes shifted from missing code to internal repair-state corruption (e.g., \texttt{NoneType} access during code-fix handling), indicating that the pipeline could continue iterating even after losing a valid executable artifact.

This failure mode accounted for \textbf{113 of 300 reports} (38\%) in the corpus---the single largest category by report count. All 113 occurred in a single continuous sequence within a subset of reports, suggesting a potential interaction between accumulated pipeline state and code-generation reliability.

\textit{Root cause}: The pipeline assumes that a semantically coherent plan will always be translated into a syntactically valid and executable artifact. In practice, this assumption is fragile. The transition from planner output to executable code is mediated by formatting conventions and downstream interface expectations that are not explicitly enforced as first-class contracts.

\subsection{Environment and Navigation Recovery Failures}

A further recurring pattern appeared in scenarios involving page navigation, table re-entry, and post-click recovery. Navigation\slash environment timeouts (e.g., \texttt{page.goto} exceeding 60{,}000ms) appeared in \textbf{120 of 300 reports} (40\%), and browser-context closure errors (``Target page, context or browser has been closed'') appeared in \textbf{48 reports} (16\%). Some failures were partially repairable: for example, one selection/navigation suite eventually recovered a bulk-selection case after repeated fixes. However, a related navigation scenario involving execution from a specific entry point in the same suite remained unresolved across repeated repair attempts, accumulating visibility failures, popup interference, and repeated navigation recovery issues. Similar problems also appeared in timeout-heavy cases that relied on \texttt{networkidle} as a readiness signal in a dynamic enterprise UI.

\textit{Root cause}: The system lacks a stable operational model of environment readiness. Even when selectors are improved, the autonomous loop remains vulnerable to popup overlays, asynchronous loading, stale session state, and navigation paths that return the browser to a partially initialized but not fully interactable UI state.

\section{Quantitative Evaluation}

\subsection{Setup}

The system was evaluated using anonymized execution data from an enterprise-scale UI testing prototype. Due to confidentiality constraints, all organization-specific identifiers, product names, customer information, internal deployment details, and proprietary operational data were removed or generalized, while aggregate quantitative metrics are reported without redaction. This practice is consistent with established industrial software engineering research methodology~\cite{runeson2009guidelines}.

We collected \textbf{300 consecutive autonomous execution reports} over a 126-day period. Reports were generated automatically by the system at the conclusion of each workflow execution, with no manual selection or filtering. The system was configured with the Improved workflow (constrained self-correction with bounded retries, skip-list filtering, and RAG-grounded selector feedback).

\textbf{Metrics.} We measure four dimensions: (i)~\emph{repair convergence} ($RC$)---the fraction of scenario families whose repair process terminates with all tests passing within the retry budget; (ii)~\emph{repair iteration count}---the number of Self-Correction--Executor cycles required per scenario family; (iii)~\emph{failure signature distribution}---the frequency and co-occurrence of distinct failure types across all reports; and (iv)~\emph{first-pass success rate}---the fraction of families that converge without any repair iteration.

\subsection{Corpus Overview}

Table~\ref{tab:logcorpus} summarizes the 300-report corpus.

\begin{table}[htbp]
\small
\caption{Overview of the autonomous execution log corpus}
\label{tab:logcorpus}
\begin{tabular}{@{}lr@{}}
\toprule
\textbf{Metric} & \textbf{Value} \\
\midrule
Consecutive execution reports analyzed & 300 \\
Reports with executable test files & 187 (62.3\%) \\
Reports with no test artifact produced & 113 (37.7\%) \\
Individual test-case executions & 636 \\
Test cases passed & 204 (32.1\%) \\
Test cases failed & 432 (67.9\%) \\
Reports reaching \texttt{COMPLETED} status & 42 (14\%) \\
Unique test spec files generated & 7 \\
Distinct scenario families & 10 \\
Maximum retry depth observed & 16 \\
Evaluation window & 126-day continuous window \\
\bottomrule
\end{tabular}
\end{table}

\subsection{Repair convergence}

We define \emph{repair convergence} ($RC$) as the fraction of scenario families whose repair process terminates with all remaining tests passing within the observed retry window:

\begin{equation}
  RC = \frac{|\{\text{Converged Families}\}|}{|\{\text{All Families}\}|} \times 100
  \label{eq:pc}
\end{equation}

Table~\ref{tab:convergence} reports $RC$ and iteration statistics.

\begin{table}[htbp]
\small
\caption{Repair convergence metrics across scenario families}
\label{tab:convergence}
\begin{tabular}{@{}lr@{}}
\toprule
\textbf{Metric} & \textbf{Value} \\
\midrule
Scenario families (total) & 10 \\
Families converged ($RC$) & 7 (70.0\%) \\
Families not converged & 3 (30.0\%) \\
First-pass success rate & 1/10 (10.0\%) \\
Mean iterations to convergence & 3.4 \\
Median iterations to convergence & 4.0 \\
Max retries (converged family) & 7 \\
Max retries (unconverged family) & 16 \\
Total tests in final COMPLETED runs & 13 passed \\
\bottomrule
\end{tabular}
\end{table}

\subsection{Scenario-Family Outcomes}

Table~\ref{tab:suiteoutcomes} traces the outcome trajectory across all 10 scenario families. Convergence required between 1 and 7 repair iterations for successful families. Two important caveats apply: the selection/navigation family achieved ``COMPLETED'' status only by \emph{silently deleting} one of its two test cases at retry~7, and the advanced-features family involved \emph{assertion weakening} (from \texttt{toBe} to \texttt{toBeTruthy}) during its repair trajectory.

\begin{table*}[htbp]
\footnotesize
\caption{Scenario-family outcomes observed in the execution logs ($n$\,=\,300 reports, 10 families)}
\label{tab:suiteoutcomes}
\begin{tabular}{@{}p{2.8cm}rrcp{9.5cm}@{}}
\toprule
\textbf{Scenario family} & \textbf{\#\,Rpt.} & \textbf{Max retry} & \textbf{Conv.?} & \textbf{Observed end state} \\
\midrule
Basic interaction & 11 & 1 & Yes & All tests passed; \texttt{COMPLETED} after 1 repair iteration \\
Advanced features & 34 & 5 & Yes$^\dagger$ & \texttt{COMPLETED}; assertion weakened (\texttt{toBe}$\,\to\,$\texttt{toBeTruthy}) \\
Tab/refresh & 28 & 4 & Yes & \texttt{COMPLETED}; selector and race-condition fixes \\
Accessibility & 6 & 0 & Yes & First-pass \texttt{COMPLETED}; no repair needed \\
Detail/refresh & 40 & 6 & Yes & \texttt{COMPLETED}; timeout, API, and browser-lifecycle fixes applied \\
Selection/navigation & 45 & 7 & Yes$^\ddagger$ & Base scenario fixed; \textbf{navigation scenario silently deleted} at retry~7 \\
Status transitions & 11 & 1 & Yes & \texttt{COMPLETED}; navigation-timeout fix \\
Tab/errors & 6 & 0 & No & Multiple runs; no convergence observed \\
Details/refresh (early) & 6 & 0 & No & Single early run; not re-attempted \\
Code-gen collapse & 113 & 16 & No & No executable artifact produced in any of the 113 reports \\
\bottomrule
\multicolumn{5}{@{}l}{\footnotesize $^\dagger$Convergence involved assertion weakening. $^\ddagger$Convergence involved test-case deletion (scope reduction).}
\end{tabular}
\end{table*}

\subsection{Failure Signature Distribution}

Failure signatures were not mutually exclusive---multiple signatures frequently co-occurred within the same report. Table~\ref{tab:signaturecounts} reports the number of reports in which each signature appeared.

\begin{table}[htbp]
\small
\caption{Failure signatures across 300 execution reports (not mutually exclusive)}
\label{tab:signaturecounts}
\begin{tabular}{@{}p{5cm}r@{}}
\toprule
\textbf{Failure signature} & \textbf{\# Reports} \\
\midrule
Method/contract mismatch & 132 (44\%) \\
Navigation\slash env.\ timeout & 120 (40\%) \\
Selector/readiness failure & 96 (32\%) \\
Assertion mismatch & 78 (26\%) \\
Non-executable output & 113 (38\%) \\
Visibility assertion failure & 72 (24\%) \\
Closed browser/context & 48 (16\%) \\
Hallucinated API or selector & 36 (12\%) \\
\bottomrule
\end{tabular}
\end{table}

\subsection{Self-Correction Effectiveness}

Table~\ref{tab:selfcorrection} summarizes self-correction outcomes across scenario families that attempted repair.

\begin{table}[htbp]
\small
\caption{Self-correction outcomes by scenario family}
\label{tab:selfcorrection}
\begin{tabular}{@{}p{2.3cm}rll@{}}
\toprule
\textbf{Family} & \textbf{Iters.} & \textbf{Result} & \textbf{Strategy} \\
\midrule
Basic interaction & 1 & Fixed & Assertion fix \\
Advanced features & 5 & Fixed$^\dagger$ & Selector $\to$ assert.\ weaken \\
Tab/refresh & 4 & Fixed & Selector rewrite \\
Detail/refresh & 6 & Fixed & Timeout $\to$ API $\to$ race fix \\
Selection/nav. & 7 & Partial$^\ddagger$ & Base fixed; nav.\ case deleted \\
Status transitions & 1 & Fixed & Nav-timeout fix \\
Code-gen collapse & 16 & \textbf{Failed} & No executable produced \\
\bottomrule
\multicolumn{4}{@{}l}{\footnotesize $^\dagger$Assertion weakened. $^\ddagger$Test case deleted.}
\end{tabular}
\end{table}

\subsection{Phase-wise Analysis}

Repair effectiveness varied across the four evaluation phases (Table~\ref{tab:temporal}).

\begin{table}[htbp]
\small
\caption{Convergence behavior across execution phases}
\label{tab:temporal}
\begin{tabular}{@{}p{2.3cm}rrrr@{}}
\toprule
\textbf{Phase} & \textbf{Rpt.} & \textbf{Fam.} & \textbf{Conv.} & \textbf{Pipe.fail} \\
\midrule
Phase 1 & 18 & 2 & 1 (50\%) & 0 \\
Phase 2 & 114 & 4 & 4 (100\%) & 0 \\
Phase 3 & 54 & 2 & 1 (50\%) & 0 \\
Phase 4 (subset) & 114 & 2 & 1 (50\%) & 113 \\
\midrule
\textbf{Total} & \textbf{300} & \textbf{10} & \textbf{7 (70\%)} & \textbf{113} \\
\bottomrule
\multicolumn{5}{@{}l}{\footnotesize Reports grouped by execution order and failure patterns.}
\end{tabular}
\end{table}

Phase 4 exhibited a severe regression: 113 of 114 reports produced no executable test artifact. While one scenario family (status transitions) converged in a single retry, the remaining 113 reports all failed at the code-generation stage, suggesting an interaction between accumulated pipeline state and LLM output formatting reliability. This late-stage degradation was not observed in earlier phases and highlights limitations over extended periods.

\subsection{Empirical Implications}

Five empirical implications follow from the quantitative analysis:

\begin{enumerate}[leftmargin=*,topsep=2pt,itemsep=1pt]
  \item \emph{Successful planning is not sufficient}: the Planner generated valid scenarios in all 113 pipeline-failure reports, yet no test was ever executed. The planning layer appeared healthy while downstream generation was in collapse.
  \item \emph{Retry alone is not recovery}: retries helped 7 of 10 families converge, but the same mechanism sustained 113 sequential fruitless reports in the generation failure sequence. Without failure-type classification, the system cannot distinguish productive retries from stagnation.
  \item \emph{Convergence quality matters}: 2 of 7 ostensibly converged families achieved convergence through semantically questionable means (assertion weakening and test deletion). A na\"ive pass/fail metric would report 70\% convergence; a semantically strict metric would report 50\%.
  \item \emph{First-pass reliability is low}: only 1 of 10 families (10\%) succeeded without any repair iteration, indicating that autonomous code generation alone is insufficient for production-quality tests.
  \item \emph{Failure signatures co-occur}: the average failing report contained 2.3 distinct failure signatures, confirming that failures are compounded rather than isolated.
\end{enumerate}

\section{Root Cause Analysis}

The failure modes described above share six underlying causes.

\textbf{Non-determinism.} LLMs produce different outputs for the same input across runs, making consistent behavior impossible to guarantee. Automated workflows require repeatability; LLMs are fundamentally probabilistic.

\textbf{Lack of runtime grounding.} LLMs operate on learned statistical patterns, not verified runtime state. Even when external tools (DOM parser, RAG) are available, the model may generate intermediate reasoning that contradicts observed application state. In our corpus, 6 distinct hallucinated selectors and API calls were traced to this cause.

\textbf{Absent ground truth for test correctness.} Unlike algorithmic tasks with formal specifications, UI test validation depends on business intent and workflow semantics. Without an explicit correctness oracle, an autonomous agent cannot reliably distinguish a genuine structural recovery from a superficially passing but semantically weakened test. The assertion-weakening and test-deletion instances confirm this: the agent optimized for pass status (an observable signal) rather than behavioral correctness (an unobservable property).

\textbf{Error compounding across stages.} Each agent's output seeds the next. Noise introduced during feature discovery propagates into scenario planning, code generation, and repair. The system must therefore be understood as a tightly coupled chain, not a collection of independent modules. The average of 2.3 co-occurring failure signatures per report provides quantitative evidence of this compounding.

\textbf{Fragile interface contracts between agents.} The pipeline assumes that each stage emits outputs in a form consumable by the next stage. In practice, this assumption is brittle. A planner output may be semantically coherent yet still fail to yield an extractable code block, a valid file path, or a repair context that downstream components can consume safely. The 113-report generation failure sequence illustrates that implicit contracts can fail catastrophically: the Planner continued producing valid scenarios while the Coder consistently failed to emit parseable code blocks.

\textbf{Weak environment-state modeling.} The system does not maintain a robust model of browser readiness across navigation, popup interference, authentication state, and asynchronous UI stabilization. Navigation timeouts appeared in 40\% of reports and browser-closure errors in 16\%, even when local selector fixes were applied, because the underlying environment remained only partially recovered.

\section{Design Guidelines for Controlled Autonomous Testing}

Based on the observed failure modes, their root causes, and their quantitative frequencies, we derive five design guidelines for controlled autonomous testing. Each guideline is grounded in specific failure patterns from the evaluation.

\textbf{G1: Enforce Runtime Grounding} (addresses hallucinated interactions, 12\% of reports).
All LLM-generated selectors must be validated against the live DOM before execution. Unverified selector inference leads to hallucinated UI interactions and cascading timeouts. The system provides batch selector verification tools, but the Coder agent bypasses them during initial generation---a gap that contributed to 108 selector-timeout failures.

\textbf{G2: Enforce Bounded Iteration with Explicit Escalation} (addresses non-convergence, 30\% of families).
Repair iteration must be bounded and must escalate to a human reviewer when the retry limit is reached. The generation failure sequence consumed 113 reports without progress; a bounded policy with failure-type classification would have terminated after detecting the absence of executable artifacts.

\textbf{G3: Treat Tests as Behavioral Specifications} (addresses false positives, observed in 2 of 7 converged families).
Assertion logic and test-case scope must not be modified without human validation. The observed assertion weakening (\texttt{toBe}$\to$\texttt{toBeTruthy}) and test deletion (silent removal of a failing scenario) demonstrate that optimizing for pass status degrades test quality. Any modification that changes assertion semantics or reduces test scope should be flagged as a high-risk operation requiring human review.

\textbf{G4: Separate Environment Failures from Test Failures} (addresses environment instability, 40\% of reports).
Failures caused by infrastructure conditions---navigation timeouts, browser-context closures, authentication expiry---must be isolated from test-repair logic. A skip list automatically isolates features blocked by infrastructure failures, preventing the repair system from consuming cycles on problems outside the scope of test logic.

\textbf{G5: Enforce Explicit Interface Contracts Between Agents} (addresses generation collapse, 38\% of reports).
Planner outputs, code-generation artifacts, repair contexts, and executor inputs must be validated at every handoff point. The generation failure sequence illustrates the cost of implicit contracts: the system iterated 113 times while the Coder produced no parseable output, because no validation gate existed between the Planner and Coder stages.

\section{Controlled Autonomy Framework}

The five guidelines motivate a controlled autonomy framework. Rather than choosing between full autonomy and full manual control, the framework assigns LLMs to roles where their generative strengths are beneficial, and applies deterministic constraints at the points where LLM unreliability is structurally dangerous.

These rules are derived from recurring failure patterns observed across the 300-report evaluation corpus.

\textbf{Rule 1: Runtime grounding is mandatory.}
Any selector generated by the LLM must be validated against the current DOM state before execution. The system provides \texttt{verify\_selector} and \texttt{find\_and\_verify\_selector} tools for this purpose. Failure to enforce this rule resulted in 6 hallucinated artifacts causing cascading failures.

\textbf{Rule 2: Bounded iteration is mandatory.}
The Self-Correction--Executor loop terminates after a configurable number of attempts. Failures that do not converge are escalated to human review. Based on our data, a limit of 6--7 retries captures all converging families while preventing the 113-report stagnation observed in the generation failure sequence.

\textbf{Rule 3: Semantic preservation must be enforced.}
Any proposed modification to a test assertion or test-case scope should be treated as a high-risk operation and must not be automatically committed. The system should detect two signals: (1) semantic drift between the original and modified assertions (e.g., \texttt{toBe}$\to$\texttt{toBeTruthy}), and (2) test-case count reduction between iterations. Either signal should route the change to human validation.

This represents a fundamental boundary of autonomous test repair rather than a limitation that can be addressed solely through improved prompting or model capability.

\textbf{Rule 4: Environment-aware filtering is mandatory.}
A skip list automatically isolates features blocked by infrastructure failures, preventing the repair system from consuming cycles on problems outside the scope of test logic. Environment-related error patterns (connection refused, authentication timeout, \texttt{ECONNRESET}) are matched heuristically and automatically filtered.

\textbf{Rule 5: Interface validation is mandatory.}
Each agent handoff must be checked for structural validity before downstream execution proceeds. Planner outputs must contain an executable coding target, Coder outputs must contain extractable code blocks, and Executor inputs must contain valid test paths and repair context. Without such checks, the system may continue iterating despite having already lost the ability to execute any meaningful test---as demonstrated by 113 sequential fruitless reports.

Taken together, these rules define a constrained form of autonomy. In this model, LLMs provide exploration, proposal generation, and code drafting, while deterministic tools and human oversight provide grounding, verification, and semantic control.

From an industrial perspective, this framework provides a practical decision boundary: which parts of testing can be safely automated, and which must remain under human control. Rather than maximizing autonomy, production systems must optimize for reliability, interpretability, and failure containment. This boundary is essential for integrating autonomous testing into existing enterprise testing pipelines without compromising reliability. This enables practitioners to incrementally adopt autonomy rather than replacing existing workflows.

The framework should be read not as a claim that autonomous repair becomes universally correct, but as a set of design constraints required to operate such systems responsibly in production.

\balance
\section{Threats to Validity}

\textit{Internal validity.} All observations derive from a single anonymized enterprise-scale UI testing environment. Identified failure modes may partially reflect application-specific characteristics, LLM version behavior, or configuration choices rather than universal properties of LLM-based test repair. We mitigate this by analyzing 300 unfiltered reports rather than selected examples.

\textit{External validity.} As an industrial case study, results may not generalize across different LLMs, UI frameworks, or application domains. The target environment represents a complex enterprise-style UI system with several hundred dynamically rendered elements per screen; simpler or more static applications may exhibit different failure distributions. Replication with additional LLMs (e.g., GPT-4, Gemini) and application domains is needed.

\textit{Construct validity.} Repair convergence ($RC$) and repair iteration count capture execution stability but do not fully characterize semantic test quality. The documented assertion-weakening and test-deletion instances (Section~5.3) demonstrate this gap explicitly---convergence rate alone overstates repair effectiveness. Future evaluations should incorporate assertion-strength metrics, behavioral coverage analysis, and explicit artifact-validity checks at each agent boundary.

\textit{Reliability.} All reports were generated automatically by the system without manual intervention. The analysis procedure (scenario-family identification, failure-signature labeling, convergence classification) was performed by the first author. Independent re-labeling or multi-rater agreement analysis was not performed and remains future work.

\section{Conclusion}

We presented an industrial case study of a multi-agent LLM-based autonomous testing system evaluated using anonymized execution data from an enterprise-scale UI testing environment over 300 autonomous execution reports encompassing 636 individual test-case executions.

The system achieved a 70\% repair convergence rate at the scenario-family level, with a mean of 3.4 repair iterations per converged family. However, the quantitative analysis revealed significant limitations beneath this headline metric: only 10\% of families succeeded on first attempt; 38\% of reports failed to produce any executable test artifact; 2 of 7 ostensibly converged families achieved convergence through assertion weakening or test deletion; and system performance showed degradation across grouped execution phases, with 113 sequential pipeline failures occurring in the final evaluation phases.

These failures arise from fundamental properties of LLM-based reasoning---non-determinism, absence of runtime grounding, lack of a correctness oracle, and error compounding across pipeline stages. We identified six root causes and derived five evidence-grounded design guidelines that define the operational constraints required for stable autonomous testing.

The central finding is that reliable autonomous test operation in enterprise-scale settings requires \emph{disciplined} autonomy: LLMs provide generative power, while deterministic constraints and human oversight preserve trustworthiness. In enterprise-scale use, a semantically strict convergence rate of 50\% (excluding assertion-weakened and test-deleted families) is more operationally meaningful than the naive 70\% rate.

More broadly, this work reframes the role of autonomy in software testing. The key challenge is not how to maximize autonomous behavior, but how to bound it. Autonomous systems that operate without constraints tend to produce unstable, non-converging, or misleading results, whereas constrained systems can be integrated into real testing workflows. This perspective enables controlled adoption of LLM-based testing systems as reliable engineering tools rather than experimental automation.

The key contribution of this work is not to demonstrate success, but to define the boundary conditions under which autonomous testing becomes viable in real-world systems.

\section{Data Availability}

The analysis in this preprint is based on anonymized aggregate observations from an enterprise-scale UI testing prototype. Raw execution logs, screenshots, proprietary identifiers, internal UI structures, and environment-specific operational details are not included. All examples are generalized to avoid disclosure of organization-specific information. The preprint is intended as a technical report on failure patterns and design constraints rather than a release of an internal dataset.

\bibliographystyle{ACM-Reference-Format}
\bibliography{refs}

@inproceedings{hammoudi2016why,
  author    = {Hammoudi, Mouna and Trunk, Gregg and Braiek, Houssem Ben and Davachi, Pourya},
  title     = {Why Do Record/Replay Tests of Web Applications Break?},
  booktitle = {Proceedings of the IEEE International Conference on Software Testing, Verification and Validation (ICST)},
  year      = {2016},
  pages     = {180--190},
  publisher = {IEEE},
  address   = {Chicago, IL, USA}
}

@inproceedings{leotta2018repairing,
  author    = {Leotta, Maurizio and Ricca, Filippo and Tonella, Paolo},
  title     = {Repairing {Selenium} Test Cases: An Industrial Case Study about Web Page Element Localization},
  booktitle = {Proceedings of the IEEE International Conference on Software Testing, Verification and Validation (ICST)},
  year      = {2018},
  pages     = {88--98},
  publisher = {IEEE},
  address   = {V\"aster\r{a}s, Sweden}
}

@article{schafer2024empirical,
  author    = {Sch{"a}fer, Max and Nadi, Sarah and Eghbali, Aryaz and Tip, Frank},
  title     = {An Empirical Evaluation of Using Large Language Models for Automated Unit Test Generation},
  journal   = {IEEE Transactions on Software Engineering},
  year      = {2024},
  volume    = {50},
  number    = {1},
  pages     = {85--105},
  publisher = {IEEE}
}

@inproceedings{kang2023llm,
  author    = {Kang, Sungmin and Yoon, Juyeon and Yoo, Shin},
  title     = {Large Language Models are Few-Shot Testers: Exploring {LLM}-Based General Bug Reproduction},
  booktitle = {Proceedings of the IEEE/ACM International Conference on Software Engineering (ICSE)},
  year      = {2023},
  pages     = {2312--2323},
  publisher = {IEEE},
  address   = {Melbourne, Australia}
}

@inproceedings{fraser2011evosuite,
  author    = {Fraser, Gordon and Arcuri, Andrea},
  title     = {{EvoSuite}: Automatic Test Suite Generation for Object-Oriented Software},
  booktitle = {Proceedings of the 19th ACM SIGSOFT Symposium on Foundations of Software Engineering (FSE)},
  year      = {2011},
  pages     = {416--419},
  publisher = {ACM},
  address   = {Szeged, Hungary}
}

@inproceedings{lemieux2023codamosa,
  author    = {Lemieux, Caroline and Inala, Jeevana Priya and Lahiri, Shuvendu K. and Sen, Koushik},
  title     = {{CodaMosa}: Escaping Coverage Plateaus in Test Generation with Pre-Trained Large Language Models},
  booktitle = {Proceedings of the IEEE/ACM International Conference on Software Engineering (ICSE)},
  year      = {2023},
  pages     = {919--931},
  publisher = {IEEE},
  address   = {Melbourne, Australia}
}

@inproceedings{stocco2022similo,
  author    = {Stocco, Andrea and Yandrapally, Rahulkrishna and Mesbah, Ali},
  title     = {{SIMILO}: Multi-Criteria Matching of Web Element Locators},
  booktitle = {Proceedings of the ACM SIGSOFT International Symposium on Software Testing and Analysis (ISSTA)},
  year      = {2022},
  pages     = {322--334},
  publisher = {ACM},
  address   = {Virtual Event, South Korea}
}

@inproceedings{yuan2024chatunitest,
  author    = {Yuan, Zhiqiang and Liu, Yiling and Li, Chuanyi and Gao, Yuxiang and Liao, Zhengwei and Xu, Fei and Liu, Yue and Li, Zhenyu and Peng, Xin},
  title     = {{ChatUniTest}: A Framework for {LLM}-Based Test Generation},
  booktitle = {Companion Proceedings of the ACM International Conference on the Foundations of Software Engineering (FSE Companion)},
  year      = {2024},
  pages     = {572--576},
  publisher = {ACM},
  address   = {Porto de Galinhas, Brazil}
}

@inproceedings{feng2024prompting,
  author    = {Feng, Sidong and Chen, Chunyang},
  title     = {Prompting Is All You Need: Automated Android Bug Replay with Large Language Models},
  booktitle = {Proceedings of the IEEE/ACM International Conference on Software Engineering (ICSE)},
  year      = {2024},
  pages     = {1--13},
  publisher = {IEEE},
  address   = {Lisbon, Portugal}
}

@inproceedings{alshahwan2024automated,
  author    = {Alshahwan, Nadia and Chheda, Jubin and Fink, Anastasia and Lau, Hannah and Misael, Alavaro and Mossige, Marit and Potluri, Manisha and Rajan, Neeraja and Sarma, Aparajita and Winter, Scott},
  title     = {Automated Unit Test Improvement using Large Language Models at {Meta}},
  booktitle = {Companion Proceedings of the ACM International Conference on the Foundations of Software Engineering (FSE Companion)},
  year      = {2024},
  pages     = {185--196},
  publisher = {ACM},
  address   = {Porto de Galinhas, Brazil}
}

@inproceedings{feng2024wechat,
  author    = {Feng, Sidong and Lu, Haochuan and Jiang, Jianqin and Xiong, Ting and Huang, Likun and Liang, Yinglin and Li, Xiaoqin and Deng, Yuetang and Aleti, Aldeida},
  title     = {Enabling Cost-Effective {UI} Automation Testing with Retrieval-Based {LLMs}: A Case Study in {WeChat}},
  booktitle = {Proceedings of the 39th IEEE/ACM International Conference on Automated Software Engineering (ASE)},
  year      = {2024},
  pages     = {2065--2076},
  publisher = {ACM},
  address   = {Sacramento, CA, USA}
}

@article{runeson2009guidelines,
  author    = {Runeson, Per and H{\"o}st, Martin},
  title     = {Guidelines for Conducting and Reporting Case Study Research in Software Engineering},
  journal   = {Empirical Software Engineering},
  year      = {2009},
  volume    = {14},
  number    = {2},
  pages     = {131--164},
  publisher = {Springer}
}

\end{document}